\documentstyle[epsfig]{aipproc}
\def\mass#1{${\mathrm{#1\,M}_\odot}$}
\def\chem#1#2{$\mathrm{^{#2}\kern-0.8pt#1}$}
\def\mchem#1#2{\mathrm{^{#2}\kern-0.8pt#1}}
\def\reac#1#2#3#4#5#6{$\mathrm{\,^{#2}\kern-0.8pt{#1}\,({#3}\,,{#4})\,{}^{#6}\kern-0.8pt{#5}\,}$}
\def\betap#1#2#3#4{$\mathrm{\,^{#2}\kern-0.8pt{#1}\,(\beta^+)\,{}^{#4}\kern-0.8pt{#3}\,}$}
\def\betam#1#2#3#4{$\mathrm{\,^{#2}\kern-0.8pt{#1}\,(\beta^-)\,{}^{#4}\kern-0.8pt{#3}\,}$}
\def\reacbp#1#2#3#4#5#6#7#8{$\mathrm{\,^{#2}\kern-0.8pt{#1}\,({#3}\,,{#4})\,{}^{#6}\kern-0.8pt{#5}\,(\beta^+)\,{}^{#8}\kern-0.8pt{#7}\,}$}
\def\reacbm#1#2#3#4#5#6#7#8{$\mathrm{\,^{#2}\kern-0.8pt{#1}\,({#3}\,,{#4})\,{}^{#6}\kern-0.8pt{#5}\,(\beta^-)\,{}^{#8}\kern-0.8pt{#7}\,}$}
\def\simgr{\mathbin{\;\raise1pt\hbox{$>$}\kern-8pt\lower3pt\hbox{$\sim$}\;}}
\def\simlr{\mathbin{\;\raise1pt\hbox{$<$}\kern-8pt\lower3pt\hbox{$\sim$}\;}}
\begin{document}
\title{NUCLEOSYNTHESIS IN LOW- AND INTERMEDIATE-MASS STARS: AN OVERVIEW}

\author{Nami Mowlavi}
\address{\vskip -4mm
         Geneva Observatory, CH-1290 Sauverny, Switzerland\\
         \vskip -5mm}

\maketitle
\begin{abstract}
An overview of the main phases of the evolution of low- and intermediate-mass stars
is presented, and the different types of nucleosynthesis operating from the pre-main
sequence up to and including the asymptotic giant branch phase described.
The surface abundance modifications brought by each nucleosynthesis process is
also briefly discussed.
\vskip 1mm
\noindent{\bf To appear in:} {\it Tours Symposium on Nuclear Physics III},
Ed. H. Utsunomiya, Amer. Inst. of Phys., 1997 ({\sl invited review})
\end{abstract}

\section{Introduction}

Low- and intermediate mass (LIM) stars are defined as those who end their life without
proceeding through the carbon and heavier elements burning phases.
Part of them experience only the core H-burning phase, ending
their life as He white dwarfs (WDs). This is the case for stars with initial masses
between $\sim 0.08$ and \mass{\sim 0.5} (\mass{} being the solar mass).
Stars with masses between $\sim 0.5$ and \mass{6-8}, on the other hand,
proceed further to the core He-burning phase, and
end as C-O WDs. Stars less massive than
\mass{\sim 0.08} never reach central temperatures high enough to ignite
H, and end as brown dwarfs. Stars more massive than \mass{\sim 10}
(called massive stars), on the other hand, continue their evolution through central C
and heavier element burning, eventually ending their life in a supernova explosion.
The mass limits defining these categories are obtained through evolutionary
model calculations (e.g. \cite{maeder_meynet89}). They are still subject to
some uncertainties mostly due to the shortcomings in the mixing prescriptions
and mass loss rates. As for stars with initial masses between $6-8$, their fate
is still uncertain. We refer to \cite{maeder_meynet89} for a discussion (see
also \cite{hashimoto_etal93,garcia-berro_etal97}).

\begin{figure}
\begin{center}
\epsfig{file=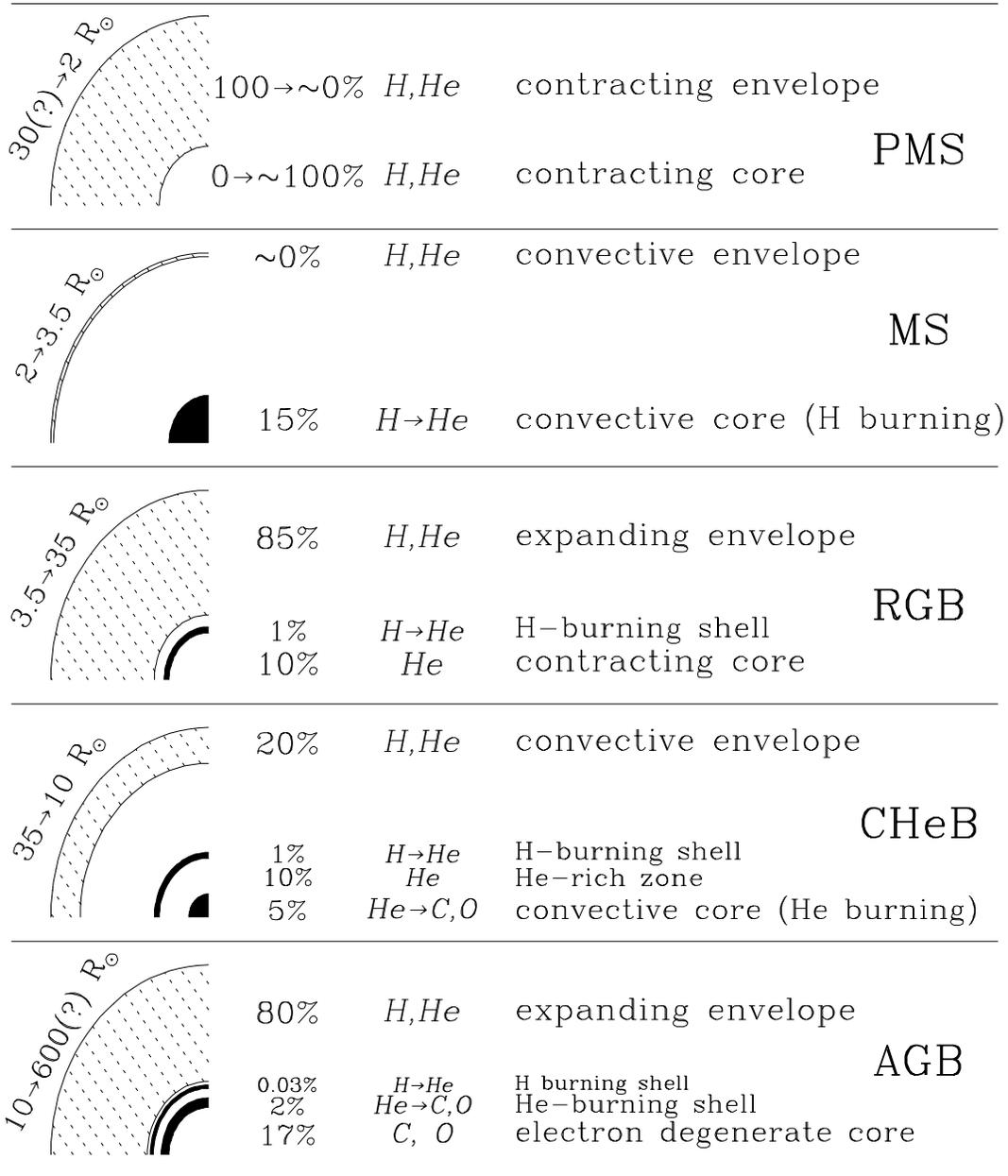}
\end{center}
\caption{\label{Fig:struct}
         Schematic representations of the structure of a star at different phases
         of its evolution. The numbers on the upper left of each diagram indicate the
         variation of the surface radius during the considered phase of evolution. The
         columns on the right of each diagram indicate, from left to right, the percent of
         the total mass of the star contained in the given region, its main chemical
         composition and a description of it. hatched areas in the diagrams indicate
         convective regions, while filled areas denote regions where nuclear energy is
         produced. The quantities refer to a \mass{3} Z=0.02 star. Adapted from
         \protect\cite{mowlavi95}.
        }
\end{figure}

In this paper, we consider the evolution of {\it single} stars with masses between
$\sim 0.5$ and \mass{6-8} (the case of massive stars is reviewed by G. Meynet in this
volume). The nucleosynthesis operating in these LIM stars
involve mainly the light elements up to Al, and the {\it s}-process elements.
Because of the high mass loss rates characterizing the late stages of the
evolution of these stars, the nuclides synthesized in their deep interior, and
dredged-up to the surface, efficiently contribute to the chemical enrichment of
the interstellar medium. Such is the case for He, \chem{Li}{7}, C, N, \chem{F}{19},
\chem{Al}{26}, and the {\it s}-process elements.

This paper aims at presenting a general overview of the nucleosynthesis occurring
in LIM stars. A more detailed review of the subject is presented in
\cite{mowlavi97a}. The main phases of the structural evolution of these stars
are presented in Sect.~\ref{Sect:structure}, while
Sect.~\ref{Sect:nucleosynthesis} analyses the nucleosynthesis operating in the
different phases of their evolution. Some concluding remarks on surface
abundance predictions in LIM
stars are discussed in Sect.~\ref{Sect:surface abundances}.

\section{Structural evolution}
\label{Sect:structure}

The structure of a \mass{3} Z=0.02 model star, with Z being the mass
fraction of all elements heavier than He (called the ``metallicity''), is
displayed in Fig.~\ref{Fig:struct} at different phases of its evolution. Five phases
are distinguished:

\vskip 2mm
\noindent 1) The {\sl pre-main sequence (PMS) phase}, which constitutes the overall
contraction phase prior to H ignition in the center. It is characterized
by an accretion phase during which mass from a circumstellar disk is accreted on
the forming star (\cite{bernasconi96} and references therein), and an increasing central
temperature until H ignites in the core at $12\simlr\mathrm{T_6}\simlr 25$ (where
$\mathrm{T_6}$ is the temperature expressed in units of $10^6$~K). The
chemical composition is homogeneous throughout the star, except for some of the
light nuclides (up to C) which already burn at temperatures of a few $10^6$~K.

\vskip 2mm
\noindent 2) The {\sl main sequence (MS) phase}, characterized by the
transformation of H to He in the core. It represents the star's longest
duration phase. The core is convective for stars with \mass{{\rm M}\simgr 1.2},
while H burns radiatively in lower mass stars.

\vskip 2mm
\noindent 3) The {\sl red giant branch (RGB) phase}, which follows the MS phase.
As a result of core contraction, the radius of the star increases and its surface temperature
decreases. Hydrogen burns now in a shell surrounding a H-depleted core.
{\it Low-mass stars} (\mass{{\rm M}\simlr 2}), defined as those containing an
electron-degenerate core at this phase of their evolution, may pass as much as
20\% of their life as red giants.  For {\it intermediate-mass stars}
(\mass{{\rm M}\simgr 2}), on the other hand, this phase is short,
generally less than 7\% of the MS lifetime.  This phase
is also characterized by the convective envelope penetrating into the deep
layers, the material of which has been affected by H-burning. The ashes of
H-burning are thus mixed to the surface. This is called the ``{\it first dredge-up}''.

\vskip 2mm
\noindent 4) The {\sl core He-burning (CHeB) phase}, characterized by the
transformation of He to C and O in a convective core surrounded by a thin
H-burning shell. This is the second longest lived phase in the life of the
star. In low-mass stars, He ignition proceeds in a degenerate
core, which leads to a thermal runaway called ``core helium flash''.

\begin{figure}
\begin{center}
\vskip -2mm
\epsfig{file=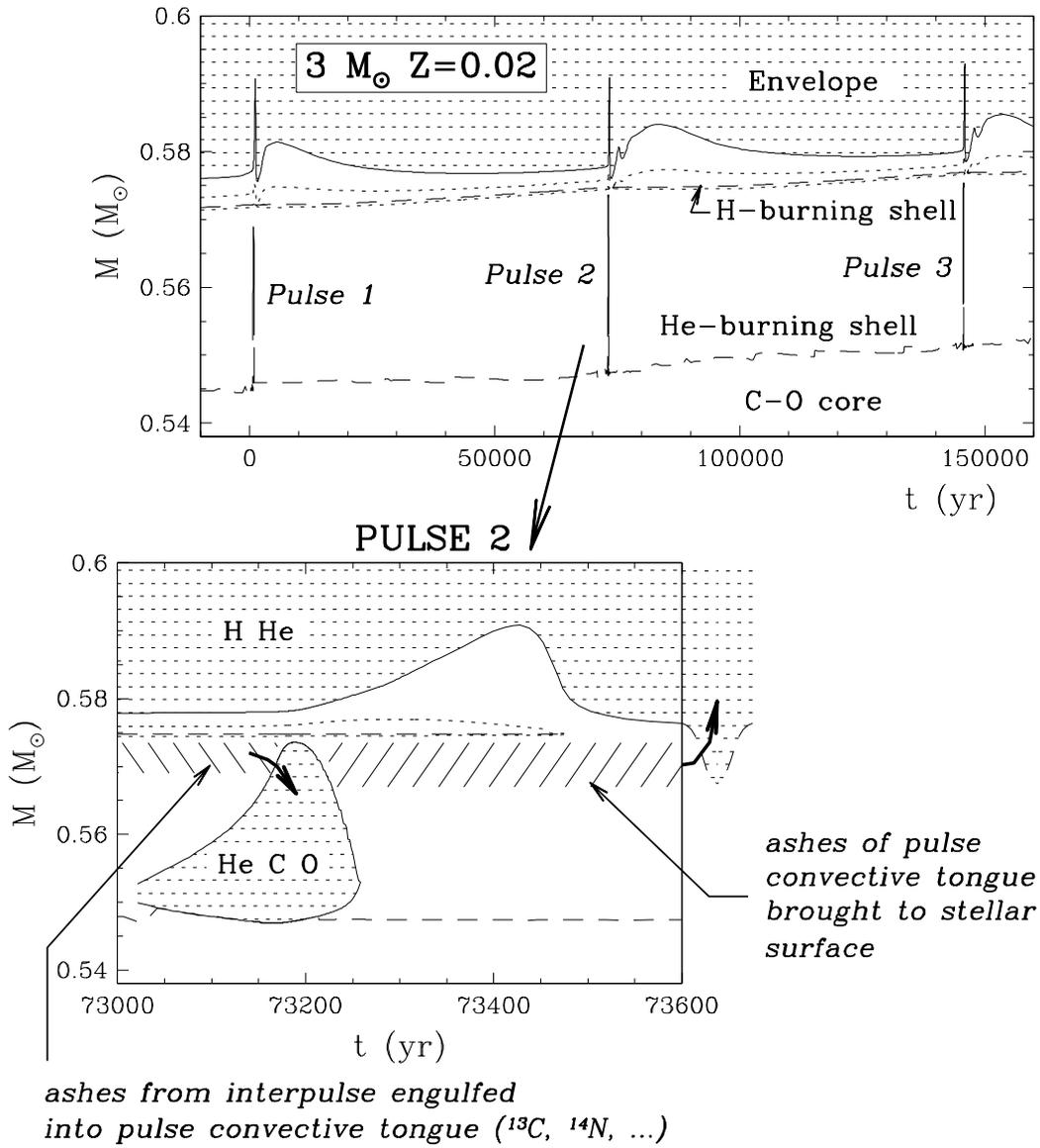}
\end{center}
\caption{\label{Fig:AGBstructure}
         {\it Upper panel:} Structural evolution of a \mass{3} Z=0.02 AGB star, during
         the first three thermal pulses in the He-burning shell. Filled areas denote
         convective regions. The long-dashed lines locate the maximum energy production
         in the H-burning (top) and He-burning (bottom) layers. Short-dashed lines locate
         the extensions of the H-burning shell, defined by the region where the energy
         production exceeds 1 erg g$^{-1}$ sec$^{-1}$. The three vertical lines in the
         He-burning shell denote the occurrence of the thermal instabilities.
         {\it Lower panel:} Enlargement of the structural evolution during the second
         pulse of the \mass{3} Z=0.02 star. A third dredge-up is simulate on the
         outer right-hand side of the panel.
         The hatched region to the left of the pulse indicates the layers containing the
         ashes left behind by the H-burning shell during the interpulse phase, and which
         are injected into the pulse.
         The layers containing the ashes of the pulse and which are mixed into the
         envelope are delimited by the hatched region on the right of the pulse.
         Adapted from \protect\cite{mowlavi95}.
        }
\end{figure}

\vskip 2mm
\noindent 5) The {\sl asymptotic giant branch (AGB) phase}, following the
CHeB phase. Helium now burns in a shell and the star becomes a red giant for the
second time.  The envelope penetrates in the deep layers and, in stars more
massive than about \mass{4}, the products of H burning are transported for the
second time to the surface (``{\it second dredge-up}''). The star is now
characterized by an electron degenerate C-O core of mass between 0.5 and
\mass{1.2}, one thin He-burning shell capped by a thin H-burning shell, and a
deep convective envelope. The nucleosynthesis and energy production are confined
to a region comprising less than 3\% of the total mass of the star.
A distinctive feature of the AGB phase is the fact that the He-burning shell
becomes thermally unstable (see Fig.~\ref{Fig:AGBstructure}) and liberates,
periodically and on a short time-scale
(several tens of years), $10^2$ to $10^6$ times the energy provided by the
H-burning shell.  These energy bursts are called ``pulses''. They lead to
the development of a convective zone in the He-burning shell. The quiescent
evolutionary phase, or ``interpulse'' period, lasts several $10^4$ years.

The lower panel of Fig.~\ref{Fig:AGBstructure} gives an enlargement of the
convective regions during a thermal instability. The thermal pulses have two
important consequences. From a chemical point of view, the material synthesized
in the He-burning shell is convectively mixed with layers close to the H-burning
shell. The layers left behind by the H-burning shell (hatched region on the left
of the pulse in Fig.~\ref{Fig:AGBstructure}), containing the ashes of that
combustion phase, are engulfed by the pulse and contribute to a rich
nucleosynthesis therein.

From a structural point of view, an important envelope response is predicted to
occur after the pulse extinction. The convective envelope penetrates into the
H-burning shell, and can even reach the H-depleted regions. Eventually,
it can sink into the carbon-rich layers. The material processed by the pulse
could then be transported to the surface. This scenario is called the
``{\it third dredge-up}''.

\section{Nucleosynthesis}
\label{Sect:nucleosynthesis}

\subsection{During the PMS phase}
\label{Sect:nucleo_PMS}

Nuclear reactions involving light elements from D to C begin to occur
in PMS stars at temperatures varying from $\mathrm{T_6}=1$ to 12.
The case of deuterium is
of particular interest in relation with Big Bang Nucleosynthesis. This element
burns at $\mathrm{T_6}\sim 1-1.5$, mainly through
\reac{D}{}{p}{\gamma}{He}{3}. As a result, the total mass of D $+$ \chem{He}{3} is
conserved. This
property, together with the fact that D is not produced in Galactic
environments, has been, and still often is, used to derive an information on
primordial D abundance (e.g. \cite{turner_etal96} and references therein).

Let us also mention the case of \chem{Li}{6} and \chem{Li}{7}, which burn
in the deep layers by
p-capture at $\mathrm{T_6}\sim 3$. To what degree the surface Li is affected
depends on the initial stellar mass and on various physical conditions such as
mixing or rotation. We refer to \cite{martin_claret96,dantona_mazzitelli94} and
references therein for a discussion on this issue.

\subsection{From the MS up to the AGB phase}
\label{Sect:nucleo_MStoAGB}

As far as stellar surface abundances and interstellar chemical enrichment are concerned,
H burning is the only nucleosynthesis process of interest in LIM
stars from the MS up to the AGB phase. Indeed, the nuclides produced by He burning
during the CHeB phase remain trapped into the white dwarfs.

Four non-explosive H-burning modes have been identified to date (e.g. \cite{rolfs_rodney88}):
the pp-chains, the ``cold'' CNO cycles, and the NeNa and MgAl
chains. The pp-chains and CNO cycles are the main contributors to the energy
production. The pp-chains are dominant at $\mathrm{T_6}\simlr 20$
(i.e. in the cores of MS stars with M$\simlr$\mass{1.2}), and the CNO cycles
at higher temperatures (i.e. in the cores of MS stars with M$\simgr$\mass{1.2}, and in
the H-burning shells). When the latter cycles are active in the central regions of MS
stars, convection develops in the core due to the steep temperature dependence of the energetics of
the CNO reactions.  All the four modes, however, are of importance from a
nucleosynthetic point of view.

The pp-chains are described at length in the literature (see \cite{clayton68,arnould_mowlavi93} for
a general description, and \cite{bahcall89,dzitko_etal95} for a discussion related
to the ``solar neutrino problem''). We discuss here only the implications
of this mode of H burning in relation with \chem{He}{3}. In addition to
its production through D burning during the PMS phase (see Sect.~\ref{Sect:nucleo_PMS}),
\chem{He}{3} is produced in low-mass stars\footnote{In stars more massive than
\mass{3-4}, the MS lifetime is shorter than the characteristic \chem{He}{3}
production time-scale, and no \chem{He}{3} is produced. In these stars,
\chem{He}{3} is essentially destroyed in the deep layers} by p(p,e$^+\nu$)\reac{D}{}{p}{\gamma}{He}{3}.
This occurs essentially in stellar regions where $\mathrm{T_6}\simlr 15$ (in
deeper layers where the temperatures are higher, \chem{He}{3} is destroyed by
\reac{He}{3}{\mchem{He}{3}}{2p}{He}{4} or \reac{He}{3}{\alpha}{\gamma}{Be}{7}),
and which are engulfed by the first dredge-up. Low-mass stars would thus be
important contributors to the galactic \chem{He}{3} enrichment. According to
current galactic chemical evolution models, however, this leads to a present
interstellar \chem{He}{3} abundance prediction much higher than what is observed
(e.g. \cite{vangioni-flam_etal94,olive_etal97,prantzos96a}). We refer to \cite{mowlavi97a} for
a more extensive discussion on this subject.

Hydrogen burning by the CNO cycles operates mainly through the
transformation of \chem{C}{12} and \chem{O}{16} to \chem{N}{14}
(\cite{clayton68}, see also G. Meynet in this volume).
We refer to \cite{arnould_etal95} for a
description of the cycles and their yields, as well as for a discussion of the uncertainties
still affecting the involved reactions rates. Two nuclides,
\chem{C}{13} and \chem{O}{17}, play an important role in confronting stellar model
predictions with observations. The nuclear transformation of \chem{C}{12} to
\chem{N}{14} proceeds first through the production of \chem{C}{13} by
\reacbp{C}{12}{p}{\gamma}{N}{13}{C}{13}, and then through its destruction by
\reac{C}{13}{p}{\gamma}{N}{14} (see Fig.~1 of \cite{arnould_etal95}). As a result,
\chem{C}{13} is overproduced in intermediate layers of LIM stars, and destroyed
in deeper regions (\cite{mowlavi97a}). The same is true for \chem{O}{17}, which is
produced by \reacbp{O}{16}{p}{\gamma}{F}{17}{O}{17} and destroyed by
\reac{O}{17}{p}{\alpha}{N}{14}. The surface abundances of both \chem{C}{13} and
\chem{O}{17} are then increased by the first dredge-up scenario. Indeed, a severe
decrease in the \chem{C}{12}/\chem{C}{13} and \chem{O}{16}/\chem{O}{17} ratios
is observed at the surface of red giant stars (e.g. \cite{eleid94}).

Arnould, Mowlavi and Champagne \cite{arnould_etal95} have also reviewed the
NeNa and MgAl chains. These become mainly active at $\mathrm{T_6}\simgr 40$. The
first of these chains contributes to the production of \chem{Na}{23} by proton
capture on \chem{Ne}{22}. The second one is of importance, prior to the AGB
phase, in the production of \chem{Al}{27} from \chem{Mg}{25} and, at $\mathrm{T_6}\simgr 65$,
from \chem{Mg}{24}. Overabundances of both \chem{Na}{23} and \chem{Al}{27} are
also confirmed at the surface of red giants (e.g.
\cite{takeda_takada-hidai94,cavallo_etal96,pilachowski_etal96}).

\subsection{During the AGB phase}
\label{Sect:nucleo_AGB}

\begin{figure}
\begin{center}
\epsfig{file=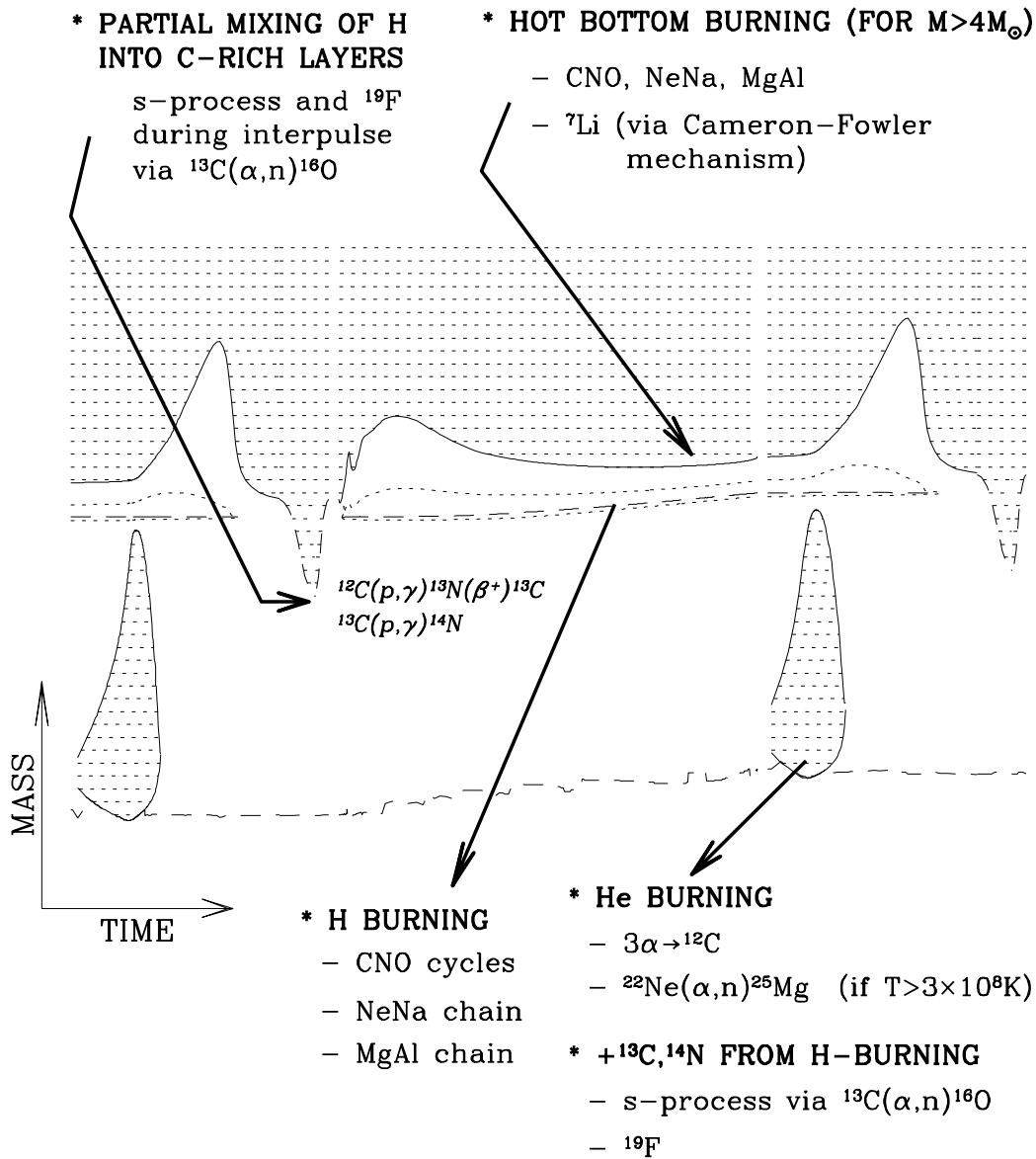,angle=00}
\end{center}
\caption{\label{Fig:AGB_nucleo}
         Overview of the nucleosynthesis occurring in AGB stars. The figure
         displays a schematic representation of two successive pulses. The interpulse
         phase is drawn on a much reduced time-scale, and lasts in reality {\it much}
         longer than each pulse. Hatched areas and dashed lines have the same meanings as
         in Fig.~\ref{Fig:AGBstructure}. Dredge-ups following each pulse (identified by
         the dashed envelope border) are also displayed. Adapted from
         \protect \cite{mowlavi95}.
        }
\end{figure}

The nucleosynthesis occurring in AGB stars is summarized in
Fig.~\ref{Fig:AGB_nucleo}. Four sites can be distinguished:

\vskip 2mm
\noindent 1) The first site is obviously the H-burning shell during the
interpulse period. The ashes of that combustion are brought to the surface
through the third dredge-up scenario. The temperature in the H-burning shell
increases with time, and reaches values above $\mathrm{T_6}$. In those
conditions, the MgAl chain leads, in particular, to an efficient production of
\chem{Al}{26} (\cite{forestini_etal91}). The interest in this nuclide resides
both in relation to $\gamma$-ray line astronomy (e.g. \cite{prantzos96b}) and
to cosmochemistry (e.g. \cite{anders_zinner93,macpherson_etal95}).

\vskip 2mm
\noindent 2) The He-burning shell constitutes the second obvious nucleosynthesis
site. It contributes mainly to the production of \chem{C}{12} through the
3-$\alpha$ process. The carbon is brought by the convective tongue of the thermal
pulses to regions close to the H-burning shell. Successive occurrences of
the third dredge-up scenario are then responsible for a gradual surface carbon
enrichment, turning eventually a M star (whose surface C/O ratio is less than 1)
into a C star (with C/O$>$1).

Another important contribution from the He-burning shell is provided by the
nucleosynthesis resulting from the injection into the pulse of the H-burning
shell ashes. In particular, the injection of \chem{C}{13}
leads to the production of neutrons
through the \reac{C}{13}{\alpha}{n}{O}{16} reaction. These neutrons can then be
used to produce elements heavier than iron via the s-process nucleosynthesis, as
well as \chem{F}{19} (the nucleosynthesis path of which is described in \cite{mowlavi_etal96}).
Unfortunately, the amount of \chem{C}{13} left behind by the H-burning shell
turns out to be too low to result in any significant s-process nucleosynthesis,
as well as to produce the fluorine in the amounts required by the observations
(\cite{mowlavi_etal96}).

\vskip 2mm
\noindent 3) An independent source of \chem{C}{13} is expected to result from a
partial mixing of protons into the carbon-rich region (\cite{herwig_etal97}). This
occurs below the convective envelope during a dredge-up scenario (see
Fig.~\ref{Fig:AGB_nucleo}). When the temperature increases during the
interpulse period, the protons are captured by \chem{C}{12} and form
\chem{C}{13}. The low proton density prevents {\it all} of this \chem{C}{13} to
be transformed into \chem{N}{14}. The surviving \chem{C}{13} is most probably burned
radiatively before the occurrence of the next pulse, leading to the production of
\chem{F}{19} and the s-process elements during the interpulse phase.

\vskip 2mm
\noindent 4) Finally, in stars more massive than about \mass{4}, the temperature
at the bottom of the envelope increases above $\mathrm{T_6}\sim 50$, activating the
H-burning modes {\it within} the envelope. This phenomenon, known as hot
bottom burning (HBB), modifies the surface composition without the need to invoke
a dredge-up scenario. The signatures of HBB are quite characteristic. In
particular, it can explain the low \chem{C}{12}/\chem{C}{13} at the surface of
some stars. It destroys \chem{C}{12} and \chem{O}{18} through the CNO cycles
(\cite{boothroyd_etal93}),
and \chem{F}{19} by \reac{F}{19}{p}{\alpha}{O}{16} (\cite{mowlavi_etal96}). It
provides also an efficient site for the production of \chem{Al}{26} through the
MgAl chain.  Moreover, HBB can lead to the synthesis of \chem{Li}{7} through the
Cameron and Fowler (\cite{cameron_fowler71}) mechanism, which combines the production of
\chem{Be}{7} at the bottom of
the convective envelope by \reac{He}{3}{\alpha}{\gamma}{Be}{7} and its convective
transport to the surface where it is transformed to \chem{Li}{7} by electron
capture (\cite{sackmann_boothroyd92}).

\section{Final Remarks}
\label{Sect:surface abundances}

  The first and second dredge-ups mix the ashes of hydrogen burning in the envelope. In
contrast, the third dredge-up has the distinctive characteristic of bringing to the surface,
in addition to the ashes from the H-burning shell, products resulting from helium burning.
Nucleosynthesis
occurring {\it in} the envelope, on the other hand, affects surface abundances
without the need to invoke a dredge-up scenario. This concerns the
light elements during the PMS phase, mainly D, \chem{He}{3} and Li, or the elements
involved in the CNO cycles and NeNa and MgAl chains in the case of HBB during
the AGB phase.

Quantitative confrontation of surface abundance predictions from stellar models
with various observations, however, reveal several disagreements
which attest of our still poor knowledge of some physical processes occurring
in the stars. The main shortcomings concern the mixing processes. Some examples
in relation to LIM stars can be found in \cite{wasserburg_etal95,charbonnel95,shetrone96}.
These are reviewed in more details in
%\cite{mowlavi97a,mowlavi97b}.
\cite{mowlavi97a}.

Finally, let us emphasize that the transition from the AGB to the post-AGB phase,
during which the envelope of the AGB star is ejected
and a planetary nebula eventually forms, is still poorly known. In
particular, the history and rates of mass loss, whether the envelope is ejected
intermittently or through a single ``super wind'' episode, and when this (these)
strong wind(s) occur during the AGB phase, all limit the predictive capabilities
of current AGB models concerning the final yields of LIM stars.

\end{document}